\documentclass[twocolumn,nofootinbib,prb,nolongbibliography]{revtex4-2}
\usepackage{amsmath,color,graphicx}
\usepackage{pifont}

\begin{document}
\title{Exploration of all-3d Heusler alloys for permanent magnets: an 
\textit{ab initio} based high-throughput study
}
\author{Madhura Marathe}
\email{madhura.marathe@physics.uu.se}
\affiliation{Department of Physics and Astronomy, Uppsala university, 
751 20 Uppsala, Sweden.}
\author{Heike C. Herper}
\affiliation{Department of Physics and Astronomy, Uppsala university, 
751 20 Uppsala, Sweden.}
\date{\today}
\begin{abstract}

Heusler alloys have attracted interest in various fields of 
functional materials since their properties can quite easily 
be tuned by composition. Here, we have investigated the relatively 
new class of all-3d Heusler alloys in view of its potential 
as permanent magnets. To identify suitable candidates, we 
performed a high-throughput study using an electronic structure 
database to search for X$_2$YZ-type Heusler systems with 
tetragonal symmetry and high magnetization. For the alloys 
which passed our selection filters, we have used a combination 
of density functional theory calculations and spin dynamics 
modelling to investigate their magnetic properties including 
the  magnetocrystalline anisotropy energy and exchange 
interactions. The candidates which fulfilled all the search 
criteria served as 
input for the investigation of the temperature dependence of 
the magnetization and determination of  Curie temperature. 
Based on our results, we suggest that Fe$_2$NiZn, Fe$_2$NiTi 
and Ni$_2$CoFe are potential candidates for permanent magnets 
with large out-of-plane magnetic anisotropy (1.23, 0.97 and 
0.82\,MJ/m$^3$ respectively) and high Curie temperatures 
lying more than 200\,K above the room temperature. 
We further show that the magnitude and direction of anisotropy 
is very sensitive to the strain by calculating the values 
of anisotropy energy for several tetragonal phases. Thus, 
application of strain can be used to tune the anisotropy 
in these compounds. 
\end{abstract}
\maketitle 

\section{Introduction}\label{sec:intro}
Discovery of new magnetic materials is vital to improve performance 
of a range of applications from data storage to renewable energy 
sources~\cite{Gutfleisch_et_al_2012}.
Permanent magnets constitute an essential component of 
electric generators used in wind turbines, and large amount of 
magnetic material is required for each 
generator~\cite{Kramer_et_al_2012,Lewis_Jimenez_2013}. 
Most commonly used permanent magnets in current devices typically 
contain rare-earth elements such as neodymium, samarium etc. 
which make these materials expensive and on top of that their mining 
is usually harmful to the environment~\cite{Bailey_et_al_2017}.  
Therefore, alternative candidates for permanent magnets are 
highly sought after to improve the overall sustainability. 
We would like to note that there is an intrinsic limit on 
the functional response of permanent magnets consisting of 
light elements determined by associated small spin-orbit 
coupling~\cite{vanVleck_1937}. However, we expect such magnets 
could replace the rare-earth magnets in mid-range applications, 
and thus reduce the overall dependence on rare earth elements. 

What makes a material a good candidate for permanent magnets? 
-- 1. its stability whether it can be synthesized in a given 
phase and structure: typically preferred symmetries are tetragonal 
or hexagonal crystal structures so that there is a single 
well-defined easy axis, 2. large magnetic moments, 3. large 
magnetocrystalline anisotropy with a preference for out-of-plane 
magnetization, and 4. Curie temperature above room temperature 
so that the ferromagnetic state is stable at working conditions 
for devices. Currently the best strategy to accelerate research 
into cost-effective and sustainable materials 
is to use the high-throughput methods, that is, 
to comb through a large number of candidates available in 
various structural databases for alloys and compounds and 
then to perform electronic structure methods calculations 
to determine the required physical quantities. 

Heusler alloys are intermetallic compounds with L2$_1$ structure 
and typically with a chemical formula X$_2$YZ. Conventionally X 
and Y elements are 3d transition metals and Z belongs to either 
group III, IV or V (main group element). 
One advantage of these alloys is the easy tuning of 
its properties obtained by varying the constituent elements, doping, 
site-disorder and/or strain making these useful for a multitude of 
functional properties from shape memory effects, half-metallicity 
to magnetocaloric response; for a general review on Heuslers see
the book edited by Felser and Hirohata~\cite{Felser_Hirohata_2015}. 
 Moreover, a number of studies have reported large 
magnetocrystalline anisotropy in Heusler alloys and have also  discussed 
possibility of its tuning via interstitial doping, strain as well as 
local ordering of atoms~\cite{Matsushita_et_al_2017,Herper_2018,Gao_et_al_2020,Sokolovskiy_et_al_2022},  
thus making Heusler alloys ideal potential candidates for permanent magnets. 

In recent years, the conventional Heusler family has expanded 
to include alloys in which all constituent elements are  
d-metals~\cite{Paula_Reis_2021}. 
We mainly focus here on magnetic properties of this novel class 
of Heusler compounds in which Z belongs to 3d metals 
termed as all-3d Heusler 
alloys~\cite{Wei_et_al_2015,Wei_et_al_2016,Ni_et_al_2018_Electronic,Ni_et_al_2019,Ozdogan_et_al_2019}. 
One of the first reports for this type of all-3d Heusler 
alloy is from Wei \textit{et al.}~\cite{Wei_et_al_2015} 
in which Ni-Mn alloys are doped with Ti to form the Heusler phase 
and additional Co-doping makes the material a strongly 
ferromagnetic shape memory alloy. 
Further studies on their functional responses have reported giant 
exchange-bias effect~\cite{Samanta_Ghosh_Mandal_2021}, giant 
barocaloric effect~\cite{Aznar_et_al_2019} as well as large magnetocaloric 
effect~\cite{Samanta_et_al_2022-JAC,Samanta_et_al_2022-PRM}. 
First-principles calculations have predicted occurrence of 
martensitic transformation in all-3d Zn$_2$YMn (Y = Fe, Co and Ni) 
alloys~\cite{Ni_et_al_2018_Electronic} as well as high-spin 
polarization in Fe$_2$CrZ and Co$_2$CrZ (Z = Sc, Ti and V) 
alloys~\cite{Ozdogan_et_al_2019}. 

In this paper, we examine this new class of materials for 
potential permanent magnets because most of the 3d metals are 
abundantly found and show strong magnetic interactions, 
and there exist well-established routes to make alloys in 
different phases from these metals especially binary alloys 
of Ni-Fe and Fe-Co and ternary alloys based on these 
alloys~\cite{Wohlfarth_1980}. 
One of the well-studied transition metal alloys is Fe-Co which 
has a high saturation magnetization, but small magnetocrystalline 
anisotropy~\cite{Wohlfarth_1980}. 
This alloy has the body-centered cubic (bcc) structure as 
the most stable phase, however first-principles 
calculations have predicted a large anisotropy for Fe-Co in the 
body-centered tetragonal 
phase~\cite{Burkert_et_al_2004,Hyodo_Kota_Sakuma_2015}. 
This observation is confirmed by experiments in which stabilization 
of the tetragonal phase was achieved by either epitaxial 
growth~\cite{Yildiz_et_al_2009} or doping with another 
element~\cite{Hasegawa_et_al_2019}. 
Typically bcc Fe-Co alloys are disordered, however 
ordered bulk or thin film conventional Heusler alloys of the 
form Fe$_2$CoZ~\cite{Du_et_al_2013,Jana_et_al_2021} and 
Co$_2$FeZ~\cite{Wurmehl_et_al_2005,Brown_et_al_2000}
(Z = Si or Ga) have been synthesized and 
reported to have large Curie temperatures. 

\begin{figure}[tb]
\centering
\includegraphics[width=0.47\textwidth]{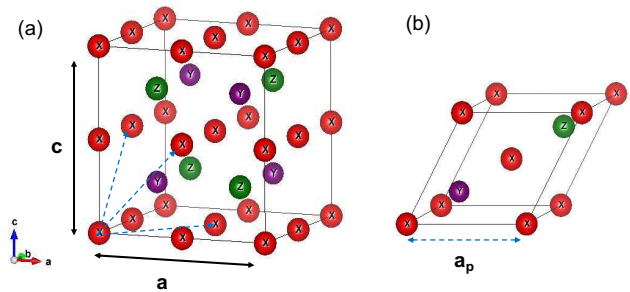}
\caption{Unit cell for a typical Heusler alloys of the type X$_2$YZ 
shown for conventional 16-atom (a) and primitive 4-atom (b). 
 Note that the two unit cells are equivalent and blue 
dashed lines 
shown in (a) correspond to the primitive cell shown in (b).  
Here filling of atomic positions corresponds to the ``standard" 
Heusler, whereas the ``inverse" structure corresponds to exchange 
of atoms  X and Y occupying positions 
($\frac{1}{2}$, $\frac{1}{2}$, $\frac{1}{2}$) and  
($\frac{1}{4}$, $\frac{1}{4}$, $\frac{1}{4}$). 
}
\label{fig:UC_comp}
\end{figure}

The main objective of our study is  
the calculation of the relevant magnetic properties 
such as magnetocrystalline anisotropy and Curie temperatures  
for a selected set of all-3d Heuslers. 
It has been shown that strain can be used to tune 
both the magnitude and the direction of easy-plane axis 
in Heusler alloys~\cite{Herper_2018,Herper_Grunebohm_2019}. 
Thus, we also explore possible tuning of the magnetocrystalline 
anisotropy for selected alloys via strain engineering. 
The details of our 
density functional theory calculations and Monte Carlo 
simulations are given in Sec.~\ref{sec:methods}. 
The selection of suitable candidates is done using 
available repository of materials and setting conditions 
on stability and magnetic moments to filter out materials 
as described in Sec.~\ref{sec:systems}. The results from 
our calculations for structural and magnetic properties 
are presented and discussed in Sec.~\ref{sec:funct}, 
then followed by analysis on how we can further tune 
the properties by strain in Sec.~\ref{sec:mae_strain}. 
Finally we conclude with a suggestion for a few potential 
permanent magnets and discuss general outlook for 
further theoretical and experimental investigations 
in Sec.~\ref{sec:summary}. 

\section{Computational details}\label{sec:methods}
We perform density functional theory (DFT) calculations using the VASP 
code~\cite{Kresse_Furthmuller_1996} to calculate the structural 
and magnetic properties of Heusler alloys. We use a 16-atom 
conventional unit cell for these calculations shown in 
Fig.~\ref{fig:UC_comp}(a). For these calculations, we use the energy 
cutoff of 650\,eV and a k-mesh of $16 \times 16 \times 16$. 
Further we use the generalized-gradient approximation of 
the Perdew-Burke-Ernzerhof (PBE) type calculated with the projected 
augmented-wave (PAW) method~\cite{Kresse_Joubert_1999}.
To calculate the magnetic anisotropy energy (MAE) and exchange 
interactions ($J_{ij}$), we make use of the full-potential 
linearized muffin-tin orbitals (FP-LMTO) code 
RSPt~\cite{Wills_et_al_2010}. To reduce the computational effort, 
we use a 4-atom primitive unit cell shown 
in Fig.~\ref{fig:UC_comp}(b) for these calculations. Further a  
dense k-mesh of $36 \times 36 \times 36$ is required to accurately 
capture smaller energy scale for magnetic interactions. 
The MAE is calculated using the magnetic force 
theorem~\cite{Wang_et_al_1996} 
and is given by:
\begin{equation}\label{eq:mae}
  \mathcal{E}_{\text{MAE}} = E_b^{[100]} - E_b^{[001]}, 
\end{equation}
where $E_b^{\alpha}$ is the fully-relativistic band energy for 
the magnetization direction $\alpha$ calculated from the self-consistent 
scalar-relativistic potential. Within this definition, a positive 
$\mathcal{E}_{\text{MAE}}$ corresponds to the uniaxial anisotropy 
and therefore, indicates a material suitable as a permanent magnet. 
The exchange parameters are calculated with the 
Liechtenstein-Katsnelson-Antropov-Gubanov formula~\cite{Liechtenstein_et_al_1987} 
as implemented in the RSPt code. 

Using the calculated $J_{ij}$'s, we map our system on a Heisenberg 
model given by spin Hamiltonian:
\begin{equation}\label{eq:HeisM}
\mathcal{H} = - \sum_{ij}J_{ij}{\bf e}_i.{\bf e}_j,
\end{equation}
where ${\bf e}_i$ is a unit vector representing local magnetic moment $m_i$ 
for site $i$. 
Note that the systems considered 
in this study contain atoms with partially filled d-electrons and 
typically have long-range exchange interactions extending over several 
lattice constants. It is essential that we include these in our model 
to correctly estimate magnetic properties at finite temperatures. 
We have included pair-wise $J_{ij}$'s for all neighbors $j$ within 
4.5$a$ radius from the central site $i$~\cite{Entel_et_al_2012}. 
We perform Monte Carlo (MC) simulations using the UppASD spin dynamics code~\cite{Eriksson_et_al-2017} to obtain finite-temperature 
properties of the system. 
Our simulation box has the dimensions  $24 \times 24 \times 24$ and 
three ensembles are used to reduce the statistical noise in our data. At each 
temperature, 50,000 steps are used to thermally equilibrate the 
system and statistical averages of physical quantities are taken 
over next 50,000 steps. A heating cycle is used to calculate 
the transition temperature with a step size of 20\,K in the 
vicinity of magnetic transition. 

\begin{figure}[tb]
    \centering
    \includegraphics[width=0.46\textwidth]{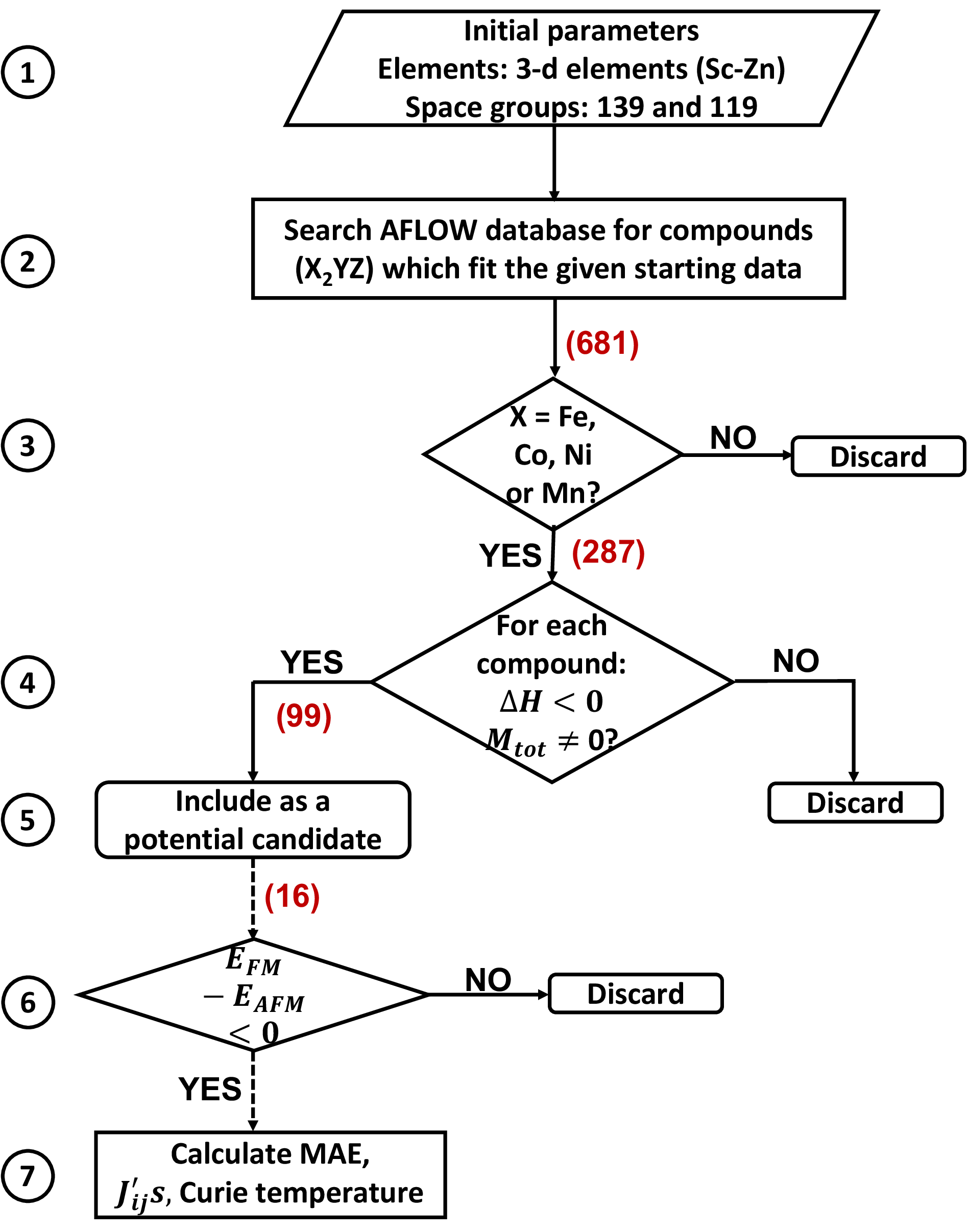}
    \caption{Flowchart describing a detailed procedure to select 
    the systems for our study (steps 1--5) using the AFLOW database 
    followed by a quick summary of DFT + MC steps for the selected 
    systems (steps 6--7). 
    The number of compounds found after each step 
    is given in brackets. For DFT calculations in step 6, we ensured 
    that there are no duplicate entries and further restricted the 
    pool to those alloys having magnetization larger than 1.0\,Tesla.
    Dashed arrows indicate that only key steps are included here. 
    }
    \label{fig:flowchart}
\end{figure}

\section{Results and discussion}\label{sec:results}

\subsection{High-throughput systems selection}\label{sec:systems}
We used the electronic structure database 
AFLOW~\cite{aflow_url,Rose_et_al_2017} 
 -- containing more than 3.5 million material entries -- 
to search for tetragonal Heusler alloys which are stable 
with negative formation of enthalpy $\Delta H$ and have 
magnetization larger than or equal to 1.0\,Tesla. We focused 
on X$_2$YZ type of alloys with X = Mn, Fe, Co or Ni and 
both Y and Z belong to 3d group from Sc to Zn, that is, 
we included only 3d transition metals in our search. 
Both standard and inverse Heusler structures (space group 
139 and 119 respectively) are included.\,\footnote{ Note that 
here the exclusion of cubic Heusler alloys 
(space groups 225 and 216) from our search criteria can be a drawback 
because even metastable tetragonal phases with negative $\Delta H$ 
will be included in our list even when the cubic phase is the most 
stable phase. However, we examine for such a possibility.
\label{fotnot:aflow} } 
A flowchart summarizing our selection procedure is depicted 
in Fig.~\ref{fig:flowchart}. 
In Table~\ref{tab:all_systems}, we have tabulated all the alloys which 
satisfied our initial filtering criteria described in  steps 1--4 
and used for DFT calculations in step 6 
in Fig.~\ref{fig:flowchart}. We also 
list corresponding lattice constant for the cubic phase and $c/a$ 
values obtained from the database. 

\begin{table}[bt]
    \centering
     \begin{tabular}{p{1.8cm} p{1cm} p{0.8cm} p{1.22cm} p{1cm} p{0.8cm} c }
    \hline
    \hline
    Compound & $a$ & $c/a$ & MO & \multicolumn{2}{c}{$M_{tot}$} & Status \\
             &  \AA  &   &   & $\mu_B$/f.u.  & Tesla &  \\
    \hline
    Mn$_2$NiTi & 5.97 & 1.48 & AFM & 0.0  & 0.0 & \ding{55} \\
    Mn$_2$TiZn & 6.01 & 1.37 & FM  & 4.73 & 1.01 & \ding{83} \\
    Fe$_2$CoNi & 5.66 & 1.33 & FM  & 7.34 & 1.88 & $\notin$ \\
    Fe$_2$CoTi & 5.83 & 1.65 & FM  & 5.18 & 1.22 &\ding{52} \\
    Fe$_2$CoV  & 5.70 & 1.51 & FM  & 4.21 & 1.06 & \ding{52} \\
    Fe$_2$MnTi & 5.81 & 1.01 & FM  & 4.16 & 1.00 & $\notin$ \\
    Fe$_2$NiSc & 6.03 & 1.63 & FM  & 5.06 & 1.08 & $\notin$ \\
    Fe$_2$NiTi & 5.85 & 1.57 & FM  & 4.50 & 1.05 & \ding{52} \\
    Fe$_2$NiZn & 5.76 & 1.47 & FM  & 5.59 & 1.36 & \ding{52} \\
    Co$_2$FeSc & 5.97 & 1.01 & FM  & 4.99 & 1.09 & $\mathsf{C}$ \\
    Co$_2$FeTi$^*$ & 5.81 & 1.59 & FM  & 4.33 & 1.03 & $\mathsf{C}$ \\
    Co$_2$FeZn & 5.70 & 1.02 & FM  & 4.91 & 1.24 & $\mathsf{C}$ \\
    Ni$_2$CoFe & 5.61 & 1.45 & FM  & 5.62 & 1.49 & \ding{52} \\
    Ni$_2$FeCu & 5.65 & 1.36 & FM  & 3.92 & 1.01 & $\notin$ \\
    Ni$_2$MnCu & 5.69 & 1.33 & AFM & 0.0 & 0.0 & \ding{55} \\
    Ni$_2$MnZn & 5.77 & 1.22 & AFM & 0.0 & 0.0 & \ding{55} \\
    \hline
    \hline
    \end{tabular}
    \caption{List of all ``eligible" X$_2$YZ alloys extracted from 
    AFLOW database including their  structural parameters obtained from 
    the database. All these compounds except Co$_2$FeTi (marked with *) 
    have the standard Heusler phase. The fourth columns lists  
    the magnetic ordering and the fifth and sixth columns list 
    the total magnetic moment in different units  
    for each compound calculated using DFT. The last 
    column gives the status of the system with regards to further 
    analysis -- 
    \ding{55}: AFM ordering; 
    \ding{83}: complex spin ordering; 
    $\mathsf{C}$: cubic structure energetically favored;
    $\notin$: small MAE values. All these criteria 
    exclude these alloys from further analysis, whereas 
    \ding{52} indicates (meta-)stable tetragonal and FM 
    phases with large MAE values suitable for full analysis
    and are discussed further in the next section. 
    }
    \label{tab:all_systems}
\end{table}

We further analyzed each of the shortlisted structures in 
the following way. First we performed DFT calculations to check 
the stability of ferromagnetic (FM) ordering versus 
antiferromagnetic (AFM) ordering (step 6 in 
Fig.~\ref{fig:flowchart}). The corresponding low-energy 
spin ordering is listed in column four of Table~\ref{tab:all_systems}. 
For the alloys found to be stable in FM ordering, we have listed 
the total magnetization $M_{tot}$ obtained from our calculations 
in columns 5 and 6 of Table~\ref{tab:all_systems}. Our values agree well 
to those reported in the AFLOW database, and all the systems 
indeed have large magnetic moments. 
We exclude those alloys for which AFM state is more stable (marked 
with \ding{55}) from further analysis.

Most conventional Heusler alloys undergo a structural phase 
transition between cubic and tetragonal phases depending on 
constituent elements and applied strain. 
We have included those alloys which have a negative 
$\Delta H$ in the tetragonal phase, but this does not ensure 
that the most stable phase is indeed tetragonal, and not 
cubic (see footnote~\ref{fotnot:aflow}). Therefore, 
for the alloys with the FM ordering, we calculated the structural 
stability against the tetragonal deformation. We keep the volume 
constant at the reference structure obtained from the database 
as we vary $c/a$ ratio from 0.8 to 1.7 for both standard and 
inverse phases (discussed in more detail for a few selected alloys 
later). From this analysis, we confirm that for X = Fe and Ni, 
tetragonal phases considered are indeed stable in the T-phase or 
have a local minima and so could be stabilized by either strain or 
depositing on a substrate. For X = Co, cubic structures are 
energetically far more favorable which is similar to what is 
observed for traditional Co-based Heusler 
alloys~\cite{Trudel_et_al_2010}. For Mn$_2$TiZn, the T-phase 
has a very shallow minima and also shows a tendency towards a more 
complex ferrimagnetic or AFM ordering with varying $c/a$ ratio. 
Therefore, it is not suitable material as a permanent magnet. 

\begin{figure}[tb]
    \centering
    \includegraphics[scale=0.35]{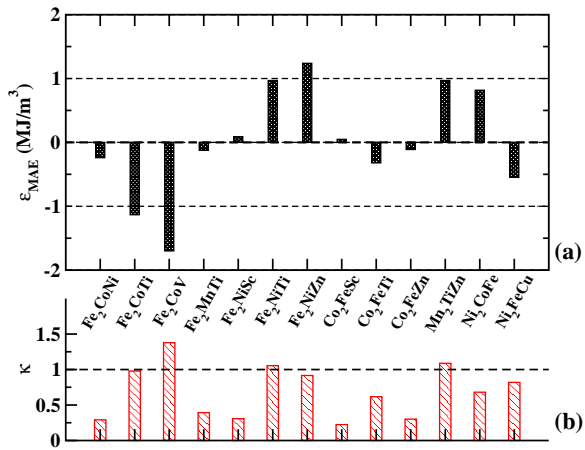}
    \caption{ 
    For all alloys which are found to be ferromagnetic (see 
    Table~\ref{tab:all_systems}) MAE values (a) calculated using 
    Eq.~\eqref{eq:mae} and magnetic hardness parameter 
    $\kappa$ (b) using Eq.~\eqref{eq:kappa} are given. 
    }
    \label{fig:mae_all}
\end{figure}

In the next step we calculated the magnetocrystalline anisotropy 
energy for 
these alloys using Eq.~\eqref{eq:mae}. The corresponding results 
are given in Fig.~\ref{fig:mae_all}(a). Furthermore using the 
calculated $\mathcal{E}_{\text{MAE}}$ and the total magnetic 
moments, we can estimate the magnetic hardness parameter 
$\kappa$ defined as~\cite{Coey_2011}: 
\begin{equation}\label{eq:kappa}
    \kappa = \sqrt{\frac{|\mathcal{E}_{\text{MAE}}|}{\mu_0M_s^2}}.
\end{equation}
Here, $\mu_0 = 4\pi \times 10^{-7}$\,JA$^{-2}$m$^{-1}$ is the 
vacuum magnetic permeability and $M_s$ is the saturation 
magnetization expressed in units of Am$^{-1}$, such that 
$\kappa$ is a dimensionless quantity. Typically $\kappa \gtrsim 1$ 
is considered as a threshold for hard magnets~\cite{Coey_2011}. 
The calculated $\kappa$ values are given in 
Fig.~\ref{fig:mae_all}(b). We find that Fe$_2$NiZn, 
Fe$_2$NiTi, Mn$_2$TiZn and Ni$_2$CoFe alloys have large and positive 
MAE values indicating a preference to have out-of-plane magnetization. 
These alloys correspondingly show $\kappa$ close to 1 with the 
exceptional case of Ni$_2$CoFe with $\kappa \sim$ 0.7 which 
results from its much larger magnetization. 
Therefore, we would further investigate all these alloys except 
Mn$_2$TiZn because of its complex spin ordering. Two alloys 
Fe$_2$CoTi and Fe$_2$CoV also have large but negative MAE which 
indicates in-plane magnetization is energetically 
favored as well as $\kappa$ close to 1. Therefore, these 
two alloys may also be interesting for applications other 
than permanent magnets and 
are included in our subsequent analysis. 
Ni$_2$FeCu alloy has a high $\kappa$ value of 0.8 but 
rather low $\mathcal{E}_{\text{MAE}}$ of $-0.55$\,MJ/m$^3$, and 
thus is not included as a potential candidate. 
All the remaining systems have much smaller MAE values of 
$\leq |0.5|$\,MJ/m$^3$ and low $\kappa$ values therefore are not 
technologically viable. 
The last column of Table~\ref{tab:all_systems} summarizes the 
final status of the each system based on our initial analysis of 
suitable candidates found in the AFLOW database. 

\subsection{Functional properties of selected alloys}\label{sec:funct}

\begin{figure*}[tb]
    \centering
    \includegraphics[scale=0.57]{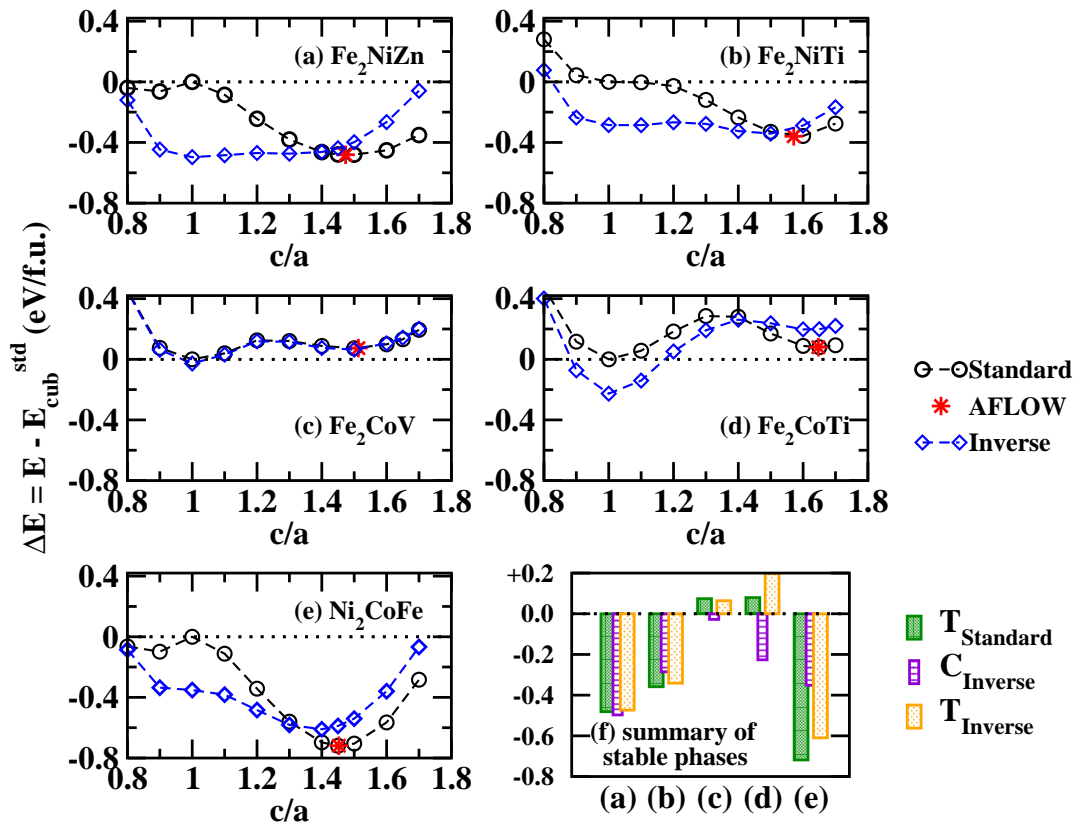}
    \caption{Stability of tetragonal phases with respect to 
    standard cubic phase: here the energy difference between 
    a distorted structure and the cubic standard phase 
    $\Delta E$ is plotted 
    as a function of $c/a$ for Fe$_2$NiZn (a), Fe$_2$NiTi (b), 
    Fe$_2$CoV (c), Fe$_2$CoTi (d) and Ni$_2$CoFe (e). 
    The results are shown for both standard and inverse phases 
    along with the structure reported in the AFLOW database. 
    (f) For easier comparison of energy scales, we summarize 
    the $\Delta E$ values for cubic and the tetragonal phases 
    at the potential well. In all cases, tetragonal standard 
    phase (green bars) corresponds to the AFLOW structure. 
    }
    \label{fig:E_cbya_x2YZ}
\end{figure*}

We first discuss in detail our structural analysis for 
tetragonal phases for the selected five alloys. 
We performed simulations for a number of tetragonal phases 
for both standard and inverse structures to determine 
minimum energy phases. These calculations are done 
such that the total volume of the unit cell is kept constant 
at its reference tetragonal structure. 
In Fig.~\ref{fig:E_cbya_x2YZ}, we have plotted the calculated 
total energy as a function of $c/a$ for the selected alloys. 
We shift the total energy with respect to that of the 
cubic standard phase of the corresponding alloy for easier 
comparison. 

For Fe$_2$NiZn and Fe$_2$NiTi alloys, 
the standard tetragonal phase is energetically favorable, but 
this phase is very close in energy to the inverse tetragonal phase 
as well as to the cubic phase. For Fe$_2$NiZn, the structure obtained 
from AFLOW (red star in Fig.~\ref{fig:E_cbya_x2YZ}) is 
lower in energy by 1.87\,meV/atom 
compared to the local minima for the tetragonal inverse at $c/a =$ 1.3 
and higher in energy by 3.99\,meV/atom as compared to the inverse 
cubic structure, whereas for Fe$_2$NiTi, the structure from AFLOW 
is the most stable structure with the corresponding energy 
differences of 4.37\,meV/atom with the inverse tetragonal phase at 
$c/a = $ 1.5 and 18.4\,meV/atom with the inverse cubic phase; 
these energy scales are clearly indicated in 
Fig.~\ref{fig:E_cbya_x2YZ}f. 
Another interesting feature of the inverse phase energy landscape 
for both the alloys is that from cubic to compressive strain minimum 
it is almost flat with very small energy 
barrier to transform from one phase to another. 
We have confirmed that the volume of inverse phase does not 
significantly vary from that of standard phase and small changes  
in volume do not affect the energy landscape. 
We note that these are zero temperature calculations, but at 
room temperature (equivalent to about 25\,meV) corresponding 
free energy landscape may differ. Moreover alloy phases will 
depend and can be controlled by synthesis conditions. 

In contrast to these two alloys 
where standard and inverse phases show distinct behaviour, Fe$_2$CoV 
alloys show almost no dependence on site occupancy by Fe atoms 
either to form standard or inverse phase as shown in 
Fig.~\ref{fig:E_cbya_x2YZ}c. This implies a complete site-disorder 
if this alloy can be formed in the Heusler phase, 
but it is more likely that it would form body-centered tetragonal 
phase as was observed in experiments~\cite{Takahashi_et_al_2018}. 
Also overall the cubic phase is more stable, the tetragonal phase occurs 
as a local minimum for this alloy. For Fe$_2$CoTi 
(Fig.~\ref{fig:E_cbya_x2YZ}d), 
the standard T-phase is in metastable state 
compared to the cubic phase and the inverse cubic phase is 
overall energetically favorable state. 
Despite their meta-stable structures and in-plane magnetization, 
we consider both Fe$_2$CoZ alloys due to their high MAE values 
and also because it has been shown experimentally that doping 
can be used to control the sign of MAE for V-doped Fe-Co 
alloys~\cite{Hasegawa_et_al_2019}. For Ni$_2$CoFe (see 
Fig.~\ref{fig:E_cbya_x2YZ}e), the tetragonal phase from AFLOW 
is indeed the most stable phase observed with the inverse tetragonal 
phase at $c/a = $ 1.4 lying 27.2\,meV/atom higher in energy. 

\begin{figure*}[tb]
    \centering
    \includegraphics[scale=0.65]{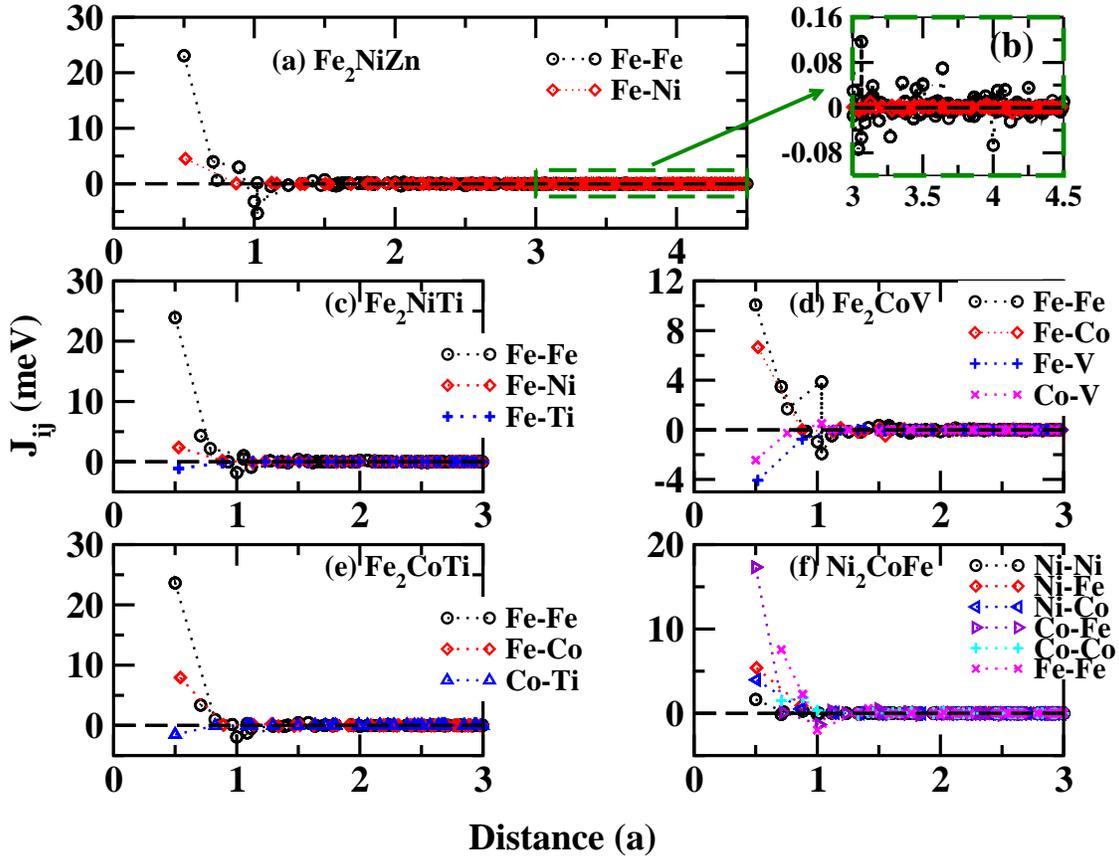} 
    \caption{Calculated pair-wise exchange interactions $J_{ij}$ for 
     Fe$_2$NiZn (a),  Fe$_2$NiTi (c), Fe$_2$CoV (d), Fe$_2$CoTi (e) 
     and Ni$_2$CoFe (f): it is plotted as a function of distance between 
     the sites $i$ and $j$ given in units of lattice constant of that 
     alloy. A positive $J_{ij}$ sign corresponds to the 
     ferromagnetic coupling between atoms at $i$ and $j$, whereas 
     a negative sign corresponds to the AFM coupling. Note that 
     the scale is different for y-axis in each panel. In panel (b), 
     we have zoomed in on slowly decaying interactions at larger 
     distances for Fe$_2$NiZn as an example; for other systems, 
     we have truncated the plots at 3$a$ for clarity. 
     }
    \label{fig:jij_dist}
\end{figure*}

Next, we study finite temperature magnetic properties for 
the selected five alloys in their tetragonal standard phase 
corresponding to the AFLOW structure. 
To map each system on the Heisenberg model, the
pair-wise exchange interactions $J_{ij}$ were computed 
 for all the alloys under study, see Fig.~\ref{fig:jij_dist}.
Note that we have calculated the pair-wise interactions between all 
four sites within the unit cell up to a distance of 4.5$a$ even 
though we have only shown here only those interactions 
for which the largest interaction was $\geq $ 1\,meV for brevity. 
As expected, we observe that the nearest-neighbor interactions are the largest in magnitude and positive, i.e. FM coupling. 
 Such strong nearest-neighbor interactions arise from 
the overlapping 3d orbitals.
Often it yields that larger values of $J_{ij}$ for  the nearest neighbor  result  
 in a high Curie temperature. 
However, note that the sign of the interactions changes from positive 
to negative depending on the distance which implies 
a competition between FM and AFM coupling within system 
which may impact the overall temperature dependence. 

For Z = Zn and Ti alloys, the interactions of Z with X and Y are 
typically negligible and would not contribute to the magnetic 
properties of the system. However, for Z = V 
(see Fig.~\ref{fig:jij_dist}d) 
we observe strong AFM coupling between nearest-neighbor
Fe-V and Co-V of the same order of magnitude as leading FM 
coupling between Fe-Fe and Fe-Co.  These trends can 
be understood by looking at the local atomic moments for 
each alloy tabulated in Table~\ref{tab:Matom}. 
Therefore, for a uniform 
description for all the alloys we retain $J_{ij}$'s for all 
four sites to describe each system. Also note that the Ni$_2$CoFe 
alloy differs from other alloy systems because this alloy 
contains only magnetic 3d metals and therefore results in 
all FM nearest neighbor interactions, whereas in the remaining 
alloys there is an AFM coupling between Z and X similar to that 
observed in conventional Heusler alloys.  

\begin{table}[tb]
    \centering
    \begin{tabular}{p{2cm} p{1cm} p{1cm} r }
    \hline
    Alloy & $M_X$  & $M_Y$  & $M_Z$  \\
    \hline
     Fe$_2$NiZn  &  2.59  &  0.56  & $-0.06$ \\
     Fe$_2$NiTi  &  2.32  &  0.36  & $-0.33$ \\
     Fe$_2$CoV   &  2.05  &  1.09  & $-0.79$ \\
     Fe$_2$CoTi  &  2.33  &  1.05  & $-0.35$ \\
     Ni$_2$CoFe  &  0.69  &  1.69  & 2.76 \\
     \hline
    \end{tabular}
    \caption{Local moments on different atomic sites for the selected 
    X$_2$YZ alloys are tabulated in $\mu_B$. Note that the two sites 
    occupied by atoms X for these alloys are equivalent and have the 
    same moment. Similar to the conventional Heusler alloys the first 
    four alloys Z atom (Ti, V or Zn) has induced moment with opposite sign  
    which is also reflected in $J_{ij}$'s shown in Fig.~\ref{fig:jij_dist}. 
    }
    \label{tab:Matom}
\end{table}

The magnitude of $J_{ij}$ decreases as the distance between the 
sites increases for all $i$ and $j$ as is expected. 
However, the reduction in magnitude is rather slow and even at 
distances of around 3$a$ (corresponding to about 15\,\AA), we 
observe significant non-zero $J_{ij}$ values. 
Use of a dense k-mesh ensures that we do not see any noise.
As an example, in panel (b) we have zoomed in on the tail part 
of $J_{ij}$'s for Fe$_2$NiZn between 3-4.5$a$. Note 
that the values are of the order of 0.1\,meV and vary between 
FM and AFM coupling. 
 It has been reported that these oscillating long-range 
interactions originate from Ruderman-Kittel-Kasuya-Yoshida interactions 
mediated via conduction electrons~\cite{Sasioglu_et_al_2008}. 
Similar behaviour is observed for all the 
system considered here (not shown) and exclusion of these large 
distance interactions can impact simulations of finite temperature 
properties as discussed later. Similar long tails for 
exchange interactions have been reported for conventional 
Heusler alloys as well~\cite{Entel_et_al_2012}. 

\begin{figure}[tb]
    \centering
    \includegraphics[scale=0.4]{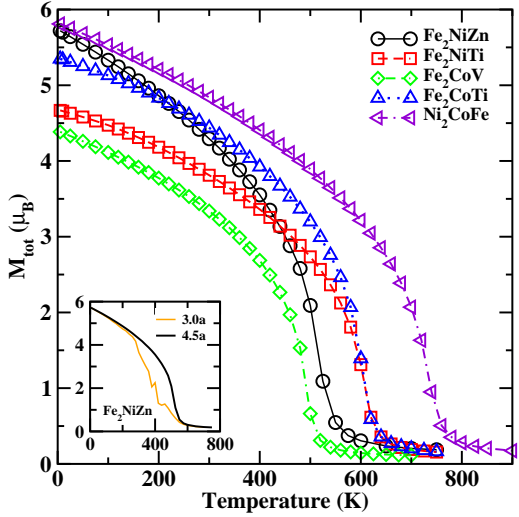}
    \caption{Curie temperature calculation: total magnetic 
    moment as a function of temperature obtained from Monte Carlo 
    simulations is plotted for different alloys. The inset compares 
    the corresponding data for Fe$_2$NiZn calculated by including 
    only short-range $J_{ij}$ interactions within the model. 
    }
    \label{fig:Tc_all}
\end{figure}

To obtain the finite temperature properties of the alloys we 
fed the $J_{ij}$'s to the Heisenberg model given in 
Eq.~\eqref{eq:HeisM}. The resulting total magnetic moments 
 of the system
$M_{tot}$ as a function of temperature are plotted in 
Fig.~\ref{fig:Tc_all}. As expected 
$M_{tot}$ decreases slowly as temperature increases 
 due to random orientations of magnetic moment of each 
individual atom at different sites arising from thermal fluctuations. 
As a result,  the material 
undergoes a second-order phase transition to a paramagnetic 
phase at high temperatures. Note that for the alloys under study 
magnetic transition occurs at much higher temperature than the 
room temperature. The calculated Curie temperature $T_c$ which 
corresponds to the peak in susceptibility from magnetization curves 
(not shown here) is highest for Ni$_2$CoFe at 740\,K followed by 
Fe$_2$NiTi (620\,K) and Fe$_2$CoTi (600\,K) and then 
Fe$_2$NiZn (520\,K) and Fe$_2$CoV (500\,K). 

Based solely on the largest observed FM interaction, we would 
expect that the largest $T_c$ for Fe$_2$NiTi, Fe$_2$NiZn or 
Fe$_2$CoTi. However, this strong FM coupling is counterbalanced 
by AFM coupling among next nearest neighbors and the largest 
$T_c$ is observed Ni$_2$CoFe for which all leading coupling terms 
are FM because of presence of all magnetic metals and AFM coupling 
is much smaller. The strongest effect of such competition is observed 
for Fe$_2$CoV in which strong AFM coupling of vanadium with 
Fe and Co (Fig.~\ref{fig:jij_dist}d) reduces both Curie 
temperature and total magnetization at zero temperature. 

These results correspond to the inclusion of $J_{ij}$'s 
corresponding to all the neighboring sites within the 
distance of 4.5$a$. To establish the effect of 
long-range interactions on temperature-dependence, we have 
compared the temperature-dependence of $M_{tot}$ for Fe$_2$NiZn 
when we exclude $J_{ij}$'s for sites at a distance larger 
than 3$a$ as shown in the inset of Fig.~\ref{fig:Tc_all}. 
Note that near the transition temperature there is a lot 
of statistical noise in the data and the system does not 
undergo a smooth transition. Such noise typically implies 
that the system has competing phases and is not able to reach 
equilibrium. For MC simulations, this may arise from either 
inadequate sampling or because ``model material" does not 
describe the ``real" system completely. 
Additional tests showed that the noise is not reduced by 
increasing either the system size, simulation time or 
the number of ensembles to improve the quality of the data. 
However, including 
long-range interactions results in a smoother transition 
confirming that the long-range nature of exchange interactions 
is essential to include in the Heisenberg model for these systems. 
This noise can be explained by competing FM and AFM coupling 
present at larger distance shown in Fig~\ref{fig:jij_dist}(b). 
Therefore, we conclude that for materials containing 3d-transition 
metals it is important to check convergence of long-distance 
interactions for better predictive power of simulations. 

\subsection{Tuning of MAE}\label{sec:mae_strain}

\begin{figure}[tb]
    \centering
    \includegraphics[scale=0.35]{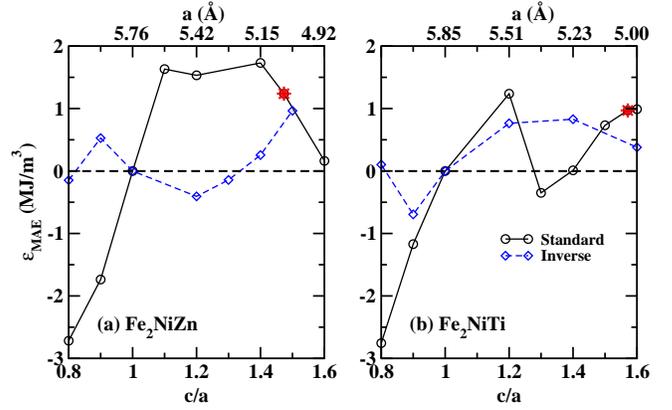}
    \caption{Effect of tetragonality on magnetic anisotropy: 
    we have plotted here the calculated MAE for Fe$_2$NiZn (a) 
    and Fe$_2$NiTi (b) as a 
    function of $c/a$ ratio for both standard and inverse phases, 
    and the corresponding in-plane lattice constants $a$ are  
    indicated at the top of the plot. 
    The AFLOW structure is highlighted with a red star and the
    corresponding MAE values are those shown in Fig.~\ref{fig:mae_all}.
    } 
    \label{fig:mae_vs_cbya}
\end{figure}

As discussed at the beginning of Sec.~\ref{sec:funct}, the 
alloys studied here show two minima (either global or local) 
at the cubic phase and the tetragonal phase with certain 
compressive strain. These two structural phases and the standard and 
inverse phases arising from chemical site ordering are shown to 
have small energy differences, therefore it should be possible 
to achieve these phases by applying strain or using different 
substrates for growing thin films. 
Typically for 3d elements, the d-orbitals lying near the 
Fermi level contribute the most to MAE~\cite{Takayama_et_al_1976}, 
and shape and position of d-orbitals is quite sensitive to 
structure, strain and chemical environment for the metal, 
which in turn means that MAE too is sensitive to these changes.  
Therefore, 
next we examine the effect of tetragonality and site-occupancy 
on the MAE values for the most promising alloys: Fe$_2$NiZn and 
Fe$_2$NiTi. 
Note that even though Ni$_2$CoFe has the highest $T_c$ and is 
quite stable in its tetragonal phase compared to the cubic phase, 
its constituent elements tend to form binary alloys  
with several competing phases~\cite{Wohlfarth_1980} 
making synthesis of Ni$_2$CoFe in Heusler structure very difficult. 
 These arguments based on existing binary phase diagrams 
are further confirmed by calculation of the phonon dispersion for 
these three alloys to check for dynamic stability of each phase; 
see Supplemental Material appended at the end for details.
This alloy system therefore would require more detailed 
study based on free energy analysis 
and is not considered for strain tuning.  

The calculated MAE values as a function of $c/a$ are plotted 
in Fig.~\ref{fig:mae_vs_cbya} for both standard and inverse 
phases. In agreement with earlier observations, we observe 
that the strain leads to changes in the magnitude of MAE and 
also results in rotation of the easy plane axis from 
out-of-plane to in-plane axis for both standard and inverse phases. 
For inverse phases (blue diamonds), 
overall smaller $|\mathcal{E}_{\text{MAE}}|$ is observed for both 
strain regimes. Significantly for applications, 
in the vicinity of stable AFLOW structures (shown with red star) 
values of MAE are comparable for standard and inverse phases. 
This indicates that site disorder should not affect the MAE 
for these alloys. 

For both alloys in the standard phase, the MAE is 
negative for tensile strains implying in-plane magnetization, 
its magnitude increases with increasing strain. However, 
we observe highly non-monotonic trends in MAE as we go from  
tensile ($c/a < 1$) to compressive ($c/a > 1$) strain. 
For Fe$_2$NiZn standard phase, the tetragonal phases with $c/a$ 
between 1.1 to 1.5 have large positive MAE of $> 1$\,MJ/m$^3$ 
which is very promising for applications, whereas for Fe$_2$NiTi, 
positive MAE observed for the ground state ($c/a = 1.57$) 
reduces almost to zero at $c/a = 1.4$ before going to a large 
positive value of 1.2\,MJ/m$^3$ at $c/a = 1.2$. This trend 
implies that very accurate control of growth or preparation 
condition would be needed for potential applications for the 
latter alloy. 
For conventional Ni-based Heusler alloys, the density of 
states analysis based on occupation and position of d-orbitals 
worked well~\cite{Herper_2018}, however for the alloys studied 
here the correlation is more complex and not obvious. 
Based on the in-plane lattice constants for 
cubic and tetragonal phases (indicated as top x-axis in 
Fig.~\ref{fig:mae_vs_cbya}), we suggest a few suitable substrates 
-- GaAs having lattice constant 4.00\,\AA\ for cubic phases 
or Cu with 3.61\,\AA\ for tetragonal phases at the local minima
(see Table 1 in Ref.~\cite{Zhu_Zhao_2013}) --  
which have small lattice mismatch calculated with respect to 
$a_p = a/\sqrt{2}$ and therefore can be used to grow 
the ``correct'' phase for tuning anisotropy. 

\section{Summary and Conclusions}\label{sec:summary}
We used a combination of high-throughput database search,  
density functional theory calculations and Monte Carlo 
simulations to study a new subclass of Heusler alloys -- 
all-3d Heuslers -- as potential candidates for permanent 
magnets. 
After application of first filters with a set of criteria we 
obtained about 20 systems with tetragonal symmetry and high 
magnetic moments in the AFLOW database. For these alloys, 
we performed a thorough examination of 
preferred easy-plane axis and Curie temperatures to 
find three potential candidates -- Fe$_2$NiZn, Fe$_2$NiTi and 
Ni$_2$CoFe -- with magnetocrystalline anisotropy of the 
order of 1 MJ/m$^3$ with preferred out-of-plane magnetization 
and which remain ferromagnetic for at least 200\,K above 
the room temperature. For Fe$_2$NiZn and Fe$_2$NiTi, the energy 
landscape is rather flat implying strain can stabilize the 
tetragonal phase. We further showed for these 
alloys that strain engineering is a viable option to 
further tune their anisotropy and site disorder between 
Fe and Ni would not reduce MAE significantly. 
 A few conventional Heusler alloys show potential 
for spintronics applications, however, we obtain low spin 
polarization values for these alloys (see Supplemental 
Material appended at the end). Therefore, these three alloys do not appear 
viable for spintronics applications.

We also found two alloys -- Fe$_2$CoV and Fe$_2$CoTi -- 
with high Curie temperature and large MAE, but with in-plane 
easy axis. These alloys are good candidates for further 
studies to examine whether alloying and/or doping can be 
used to rotate the easy axis. 
Moreover some of the alloys which did not pass our initial 
filters for permanent magnets showed some promising properties 
and could be studied further for applications such as 
magnetocalorics. 

Our study opens up a new class of rare-earth free 
permanent magnets with extensive potential possible via 
thin film growth for strain tuning. We would like to note 
that due to absence of heavy metals, there is admittedly 
a limit to the largest anisotropy obtained 
within this class of Heusler alloys. However, there 
exists a large market for intermediate applications 
of permanent magnets which do not require strong anisotropy. 
We believe that usage of these magnets for such applications 
can result in overall reduction in dependence on the 
rare-earth metals. 

\section{Acknowledgements}
We acknowledge financial support from Olle Engkvist foundation,
StandUp, eSSENCE, the Swedish Strategic Research Foundation (SSF) 
(Grant EM16-0039), and the ERC. Computational resources and 
support was  provided by the Swedish National Infrastructure 
for Computing (SNIC) at NSC (Link\"oping) and PDC 
(KTH Stockholm). 
MM also thanks Patrik Thunstr\"om and Rafael Vieira for fruitful
discussions and technical support running the RSPt and 
UppASD codes. 

\bibliography{literature_x2yz}

\clearpage
\newpage

\section*{Supplementary data for
``Exploration of all-3d Heusler alloys for permanent magnets: an 
\textit{ab initio} based high-throughput study" }

For Fe$_2$NiZn, Fe$_2$NiTi and Ni$_2$CoFe -- three alloys which show 
promising magnetic properties -- we have done additional analysis.  
The following results are obtained for the stable tetragonal phase 
found from density functional theory calculations shown 
in Fig.~4 of the manuscript. 

\section*{S1. Phonon band structures}

\begin{figure*}[bt]
    \centering
    \includegraphics[scale=0.5]{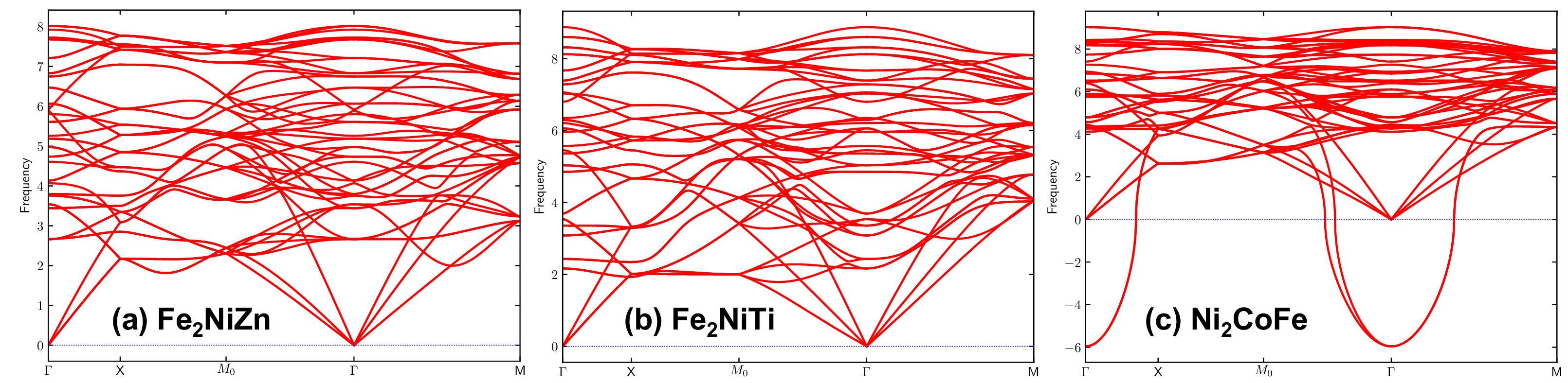}
    \caption{Calculated phonon band dispersion for the tetragonal 
    Heusler alloys Fe$_2$NiZn (a), Fe$_2$NiTi (b) and Ni$_2$CoFe (c). 
    }
    \label{fig:supp_phon}
\end{figure*}

We used PHONOPY~\cite{Togo_Tanaka_2015} code to obtain phonon 
dispersion curves 
for the tetragonal phases of three Heusler alloys. The calculations 
were performed within the harmonic approximation using the finite 
displacement method. We used a $2 \times 2 \times 2$ supercell 
constructed from a conventional 16-atom unit cell and used 
displacement of 0.01\AA\ to calculate the forces with the VASP 
code~\cite{Kresse_Furthmuller_1996}. Other computational details 
are the same those used to perform structural analysis. 

The calculated phonon frequencies along the high-symmetry paths in 
the Brilloin zone are plotted in Fig.~\ref{fig:supp_phon}. 
We indeed find that both Fe$_2$NiZn and Fe$_2$NiTi alloys are 
dynamically stable with no imaginary frequencies observed in the 
phonon dispersion, whereas we get unstable phonon modes at the 
$\Gamma$ point for Ni$_2$CoFe. Unstable $\Gamma$ modes typically 
indicate that atoms prefer to displace with respect to each other. 
This is expected in the case of Ni$_2$CoFe because its constituent 
elements have very stable binary phases and the Heusler tetragonal 
phase studied here is likely to be unstable. These results confirm our 
arguments based on binary phase diagrams of Fe-Co and Fe-Ni, that 
this system would not be a suitable candidate as a permanent magnet. 
On the other hand, the dynamic stability of Fe$_2$NiZn and 
Fe$_2$NiTi corroborates that 
these two alloys are ideal permanent magnet candidates. 
Here, we would like to note that we have not relaxed the structure of 
the systems to allow for shape changes, so there is a possibility that  
the system prefers symmetry other than tetragonal as the ground 
state. However, a full analysis of all symmetries is beyond the 
scope of this paper.

\section*{S2. Density of states}
\begin{figure}[bt]
    \centering
    \includegraphics[scale=0.5]{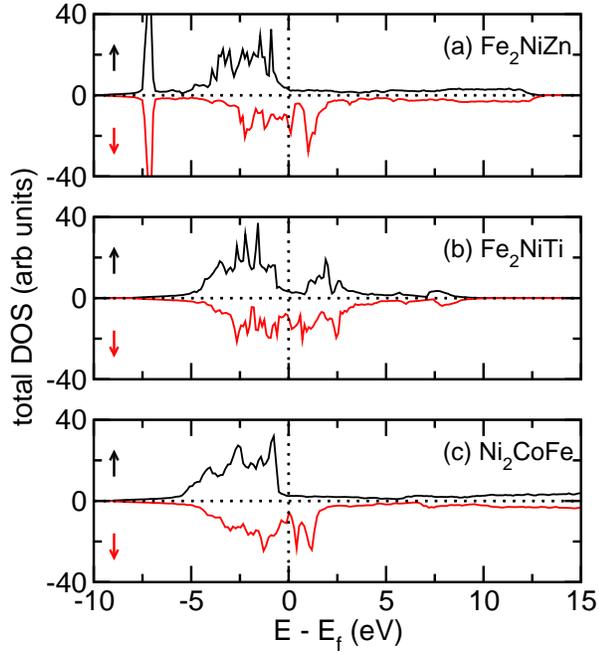}
    \caption{Spin polarized density of states calculated for the 
    tetragonal Heusler alloys Fe$_2$NiZn (a), Fe$_2$NiTi (b) and 
    Ni$_2$CoFe (c).   
    }
    \label{fig:supp_dos}
\end{figure}

For Fe$_2$NiZn, Fe$_2$NiTi and Ni$_2$CoFe, we also examined the spin 
polarized density of states shown in Fig.~\ref{fig:supp_dos}. This was 
calculated for the conventional 16-atom unit cell using the VASP package. 
For Fe$_2$NiZn and Ni$_2$CoFe, the majority spin channel is almost 
completely occupied, whereas for Fe$_2$NiTi there is a small pseudo-gap 
at the Fermi level $E_f$. This typically can result in 
large spin polarization values defined by:
\begin{equation}
    \sigma = \frac{\rho_{\uparrow} - \rho_{\downarrow}}{\rho_{\uparrow} + \rho_{\downarrow}},
\end{equation}
where $\rho_{\uparrow}$ and $\rho_{\downarrow}$ correspond to the density 
of states at Fermi level for spin up and down electrons respectively. 
The calculated $\sigma$ values for Fe$_2$NiZn, Fe$_2$NiTi and Ni$_2$CoFe 
are about 69\%, 56.6\% and 40.7\% respectively. These low values 
result from the existence of sharp peaks of localized d-states in the 
minority channel. 
Because of the existence of multiple 
localized d-states near $E_f$, small changes in the structure, strain may 
result in shifting of peaks such that the Fermi energy lies not at the 
maximum, but at a minimum, and will result in spin polarization which is 
sensitive to external parameters. For applications, materials are required 
to have robust values which can be sustained over a large operational range.
Based on these results, we conclude that these Heusler 
alloys are not half-metallic in nature and therefore would not be suitable 
for spintronics applications. 

\end{document}